\documentclass[12pt,a4paper]{article}

\usepackage{amssymb}
\usepackage{amsmath, epsfig, subfigure}

\newcommand{\be}{\begin{equation}}
\newcommand{\en}{\end{equation}}

\renewcommand{\vec}[1]{\boldsymbol{#1}}

\title{Differences in tension and compression\\in the nonlinearly elastic bending of beams}

\author{
 Michel Destrade$^a$, Jerry G. Murphy$^b$, Badar Rashid$^a$\\[0.8cm]
$^a$School of Electrical, Electronic, and Mechanical Engineering, \\
University College Dublin, Belfield, Dublin 4, Ireland;\\[0.5cm]
$^b$Department of Mechanical Engineering, \\
Dublin City University, Glasnevin, Dublin 9, Ireland.}

\date{}
\begin{document}

\numberwithin{equation}{section}

\maketitle


\begin{abstract}

The classical flexure problem of non-linear incompressible elasticity is revisited for elastic materials whose mechanical response is different in tension and compression---the so-called bimodular materials. 
The flexure problem is chosen to investigate this response since the two regions, one of tension and one of compression, can be identified easily using simple intuition. 
Two distinct problems are considered: the first is where the stress is assumed continuous across the boundary of the two regions, which assumption has a sound physical basis. The second problem considered is more speculative: it is where discontinuities of stress are allowed. 
It is shown that such discontinuities  are necessarily small for many applications, but might nonetheless provide an explanation for the damage incurred by repeated flexure.
Some experimental evidence of the possibility of bimodularity in elastomers is also presented.

\end{abstract}



\newpage


\section{Introduction}


Many materials seem to behave differently in tension and compression. 
Such materials are sometimes termed \emph{bimodular materials}. 
Such bimodularity has been observed, for example, in rocks (Lyakhovsky et al. \cite{Lyak97}), nacre (Betoldi et al. \cite{BeBD08}), and soft biological tissue, especially cartilage (Soltz and Ateshian \cite{Cart}). 
Although the concepts of tension and compression seem intuitively obvious, a precise definition of these terms is a non-trivial problem that is still open. 
Perhaps the most definitive formulation of these concepts is given in the study by Curnier et al. \cite{Swiss}, who formulate their theory in the context of unconstrained non-linear elasticity. 
The linearization of their theory yields an unambiguous and precise definition of tension and compression for infinitesimal strains, but the situation for non-linear deformations is less clear-cut, with the regions of tension and compression corresponding to positive and negative values respectively of an undefined functional of the Green-Lagrange strain tensor.

This difficulty is avoided here by considering the problem of \emph{flexure}, where intuition suggests the appropriate regions of tension and compression. 
If a rectangular bar is bent by terminal couples,  with the faces along the length of the bar assumed to be traction-free, then  the \emph{region of tension} corresponds to the region where imaginary fibres originally aligned along the length are extended and the \emph{region of compression} is where the fibres are contracted, as illustrated in Figure \ref{fig_bending} below.  
This problem is, perhaps, the benchmark problem that all theories of bimodularity must describe; specifically, a general theory of when a material is in tension and compression should coincide with our intuitive notion of tension and compression in flexure. 

The problem of flexure is considered here within the context of non-linear, isotropic, \emph{in}compressible elasticity, primarily because the mathematical analysis is considerably simpler than that for unconstrained materials. 
First formulated and solved by Rivlin \cite{Rivl49}, it has since been studied extensively in the literature (see, for example, Green and Zerna \cite{GrZe54} and Ogden \cite{Ogde84}). 
There is continuing theoretical interest in this problem, as can be seen, for example, in the recent studies of Kanner and Horgan \cite{KaHo08} and Destrade et al. \cite{DGM}. 
Rivlin \cite{Rivl49} showed that if a circular, annular sector is assumed for the deformed configuration, then an elegant solution to a natural boundary value problem can be found. 
We recall this derivation in Sections \ref{Bending_deformation} and \ref{Bending_stress}.
There, the solution is simplified  by assuming that the turning angle $\alpha$ through which the beam is bent is specified, as opposed to the more usual assumption of specifying the moment. 

\begin{figure}
\center
\epsfig{figure=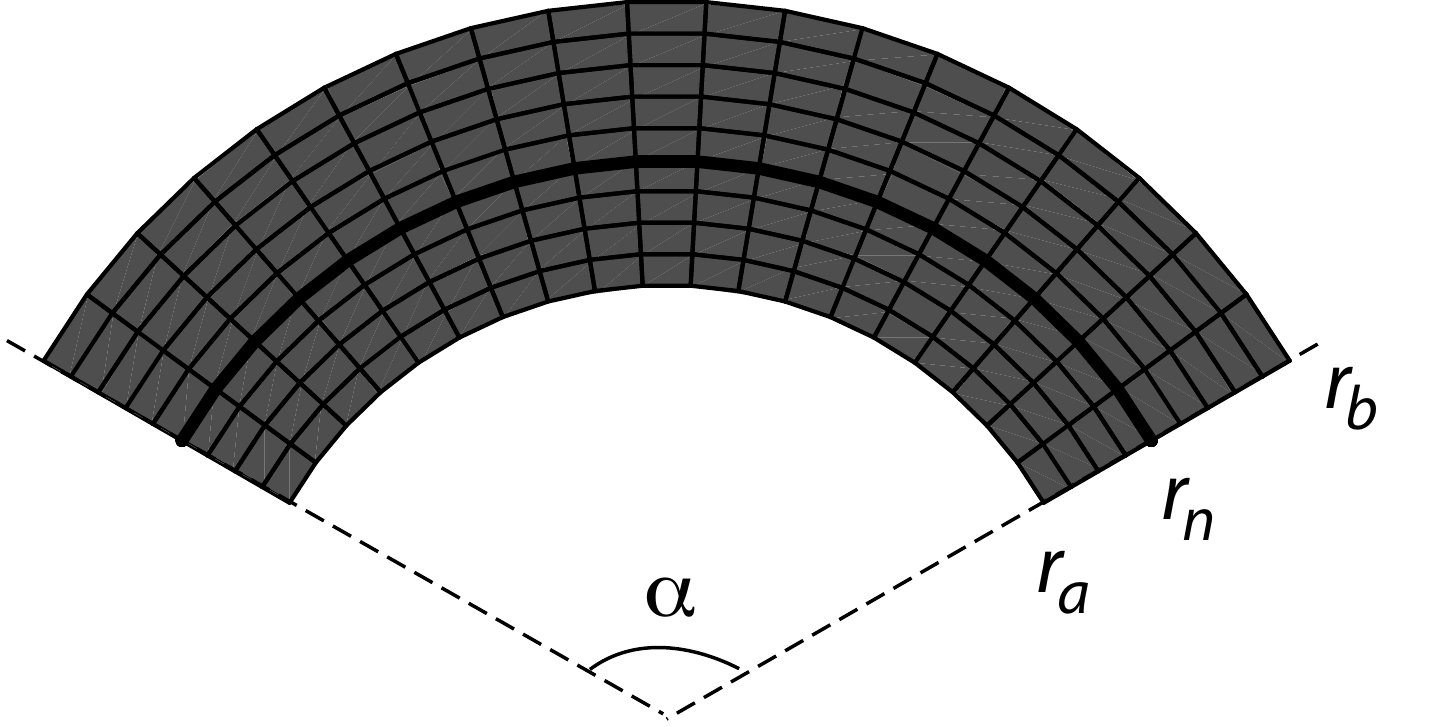, width=.8\textwidth}
 \caption{In-plane section of a bent block of an isotropic, incompressible, non-linearly elastic solid.
 The bending angle is $\alpha$. 
 The inner and outer radii of the bent faces are $r_a$ and $r_b$, respectively. 
 Along the arc line at $r=r_n$, the circumferential line elements preserve their length during bending. 
 In the region of ``compression'', $r_a \le r \le r_n$, they are contracted.
 In the region of ``tension'', $r_n \le r \le r_b$, they are extended.
 }
 \label{fig_bending}
\end{figure}

Next (\S \ref{Neutral_axis}), we establish the general expression for the stress difference across the \emph{neutral axis}, which separates the region of ``compression'' from the region of ``tension''. The consequences of assuming that the stress is continuous across the neutral axis are first explored. It is shown in Section \ref{example_section} that for the case of a Mooney-Rivlin solid, or equivalently, of a general incompressible solid in the third-order approximation of non-linear elasticity, knowledge of $\hat \mu$, the ratio of the shear moduli in tension and compression, is all that is required to solve the boundary value problem in its entirety. 
This solution is then used to illustrate some possible consequences of bimodularity in flexure. 
The possibility of a stress jump across the neutral axis is also briefly considered with an estimate given as to its possible magnitude (\S \ref{Stress_discontinuity})

The problem considered here is one of the simplest non-linear problems in solids that involve structural changes induced by applied mechanical forces. 
It is hoped that this analysis can give insight into more complicated problems involving such changes and also can provide a reference solution for non-linear flexure problems where structural changes other than simple changes in material constants are considered. 
The problem of flexure is studied as it is very often the dominant mode of deformation, in the sense that the applied forces encountered are small compared to those necessary to effect other deformations, for many materials that experience structural changes, as is shown, for example, in the work of Lua et al. \cite{sma1}. 
It is also of importance in the theoretical investigation of the consequences of proposed models of complex behaviour in solids (see, for example, Rajagopal et al. \cite{wine1}).      

Although the idea of elastomers behaving differently in tension and compression has an intuitive appeal, experimental evidence of this is scant, presumably because the effect of bimodularity has not previously been considered important. 
Some indirect evidence of this effect in rubbers is presented next.


\section{Experimental data on bimodularity}
\label{experimental_section}


Rubbers are traditionally modeled as being non-linearly elastic, incompressible materials. 
The recent data of Bechir et al. \cite{Bechir} suggest that, for at least some natural rubbers, there seems to a different response in tension and compression. 

By examining closely the data displayed on their Figure 8(a), we can estimate the slope of the stress-strain curve in the linear region of \emph{compression}. 
We obtain an excellent (visual) fit for the first recorded five data points by taking a straight line passing through the origin with slope approximatively equal to 8.33 MPa.
For the region of \emph{tension}, we use their experimental data (kindly provided by H. Bechir), see Table 1.
Here, depending  on how many points we estimate to be in the linear regime, we find that the slope is somewhere between 3.94 and 5.84 MPa.
This suggests that for the natural rubber NR70 of Bechir et al. \cite{Bechir}, the ratio of $E^+$ to $E^-$, the infinitesimal Young moduli in tension and in compression, is within the range
\be
0.47 < E^+/E^-<0.70.
\en

\begin{center}
Table 1.
\emph{Experimental data of Bechir et al. \cite{Bechir} in the early (linear) stages of the uniaxial tensile region: Cauchy stress} (MPa) \emph{Vs elongation}
(m/s).
\\[4pt]
\noindent
\begin{tabular}{l | c c c c}
\hline 
\rule[-2mm]{0mm}{8mm} 
$e$ & 0.0514 & 0.1016 & 0.1518 & 0.2023
\\
\hline
\\[-2mm]
$\sigma$ & 0.3000  & 0.4899  & 0.6525 & 0.7971
\\
 \hline
\end{tabular}
\end{center}

\begin{figure}
\centering \mbox{\subfigure{\epsfig{figure=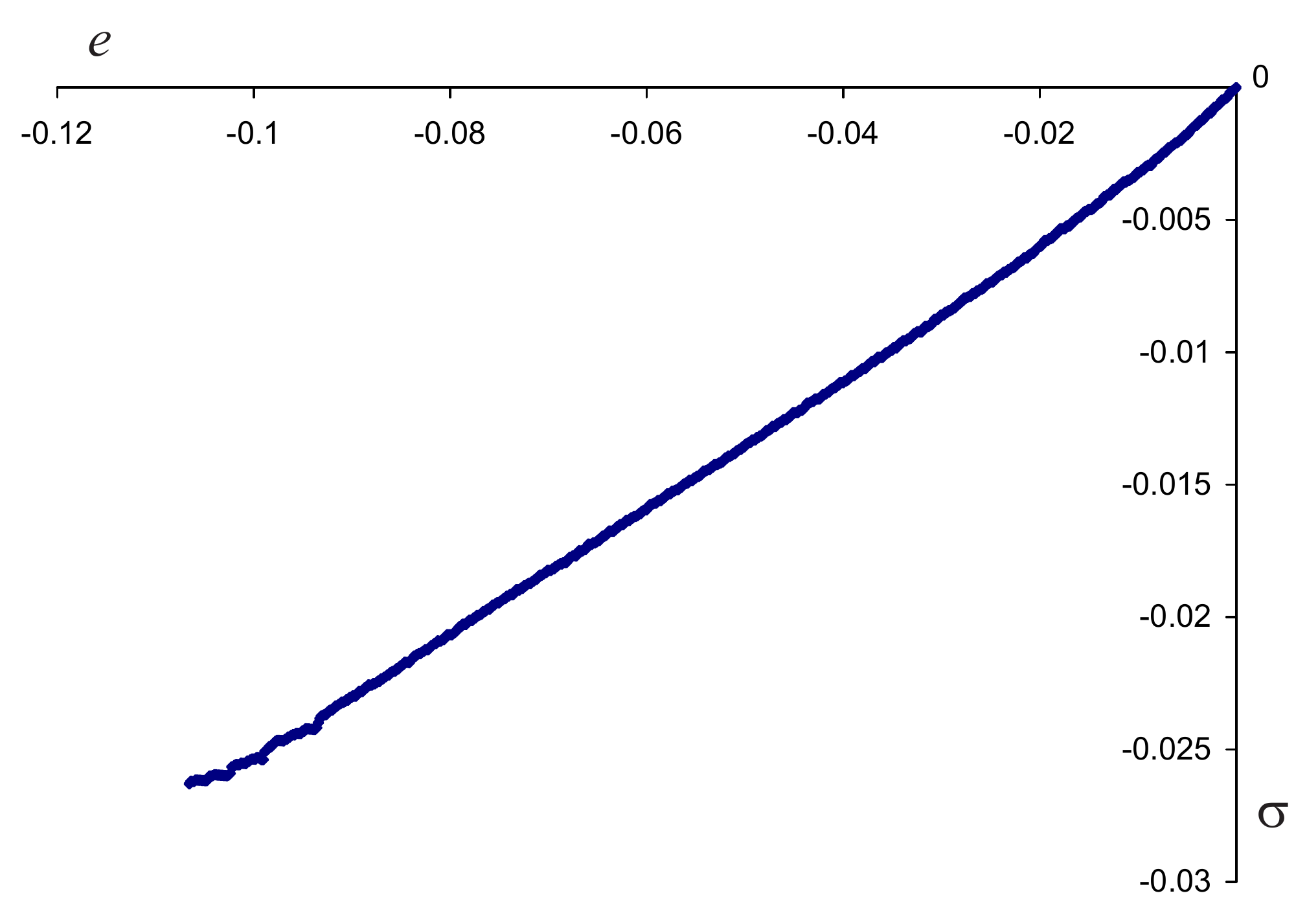, width=.45\textwidth}}}
  \quad \quad
     \subfigure{\epsfig{figure=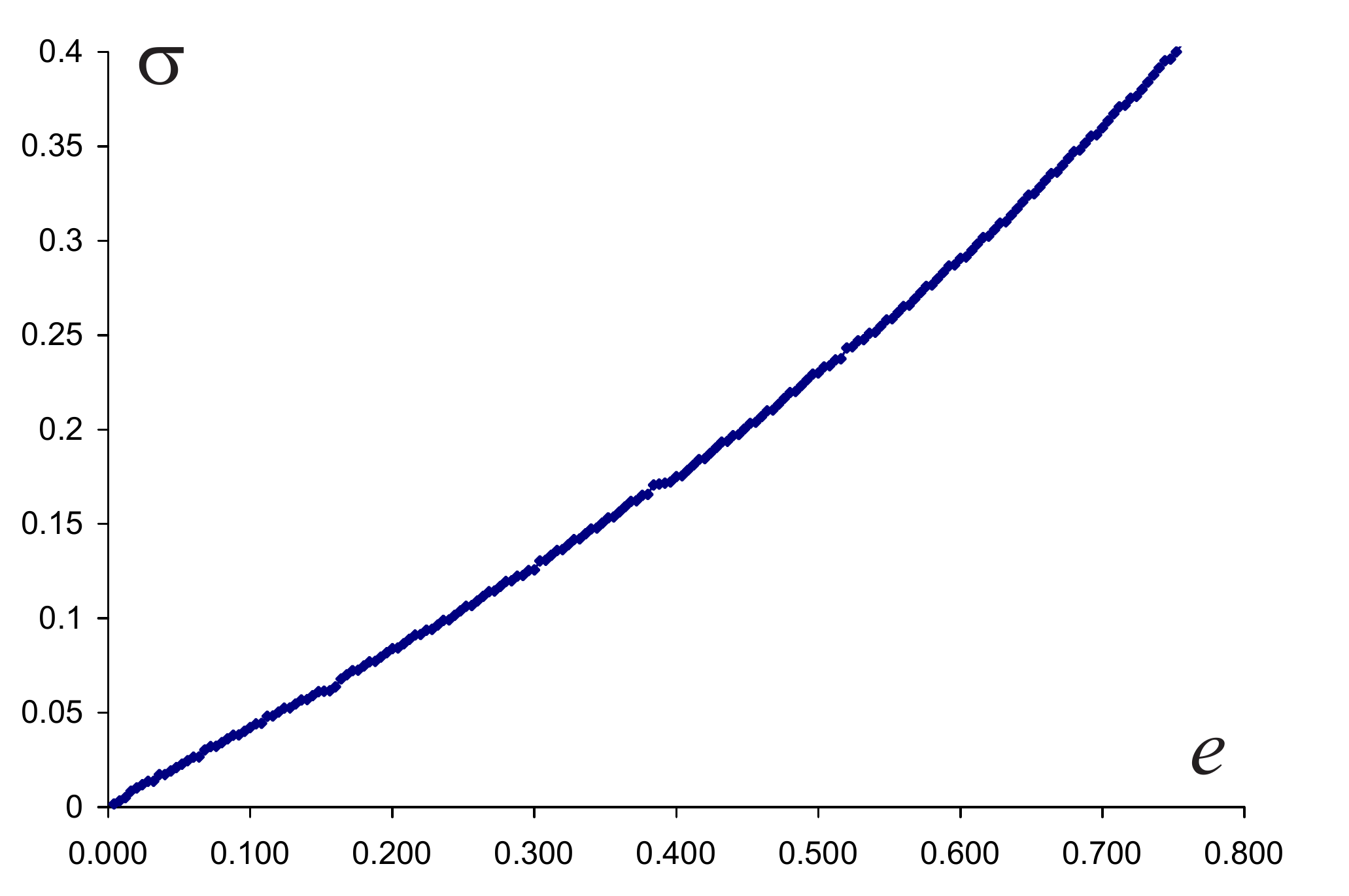, width=.45\textwidth}}
 \caption{Uni-axial compression and tension tests of translucent silicone. 
 The original dimensions of the block in compression were $50$ mm (depth) $\times$ 70 mm (width) $\times$ 50 mm (height); there, the elongation was measured as (recorded displacement)/(original height); the Cauchy stress was computed from the recorded nominal force applied by the Tinius Olsen machine; the contact areas between the compression platens and the specimen were generously lubricated.
 The dogbone specimen used in tension had an original section area of 2 mm $\times$ 6 mm; there elongation was tracked directly by LASER monitoring.}
 \label{tests}
\end{figure}

We conducted similar experiments on soft translucent silicone [Feguramed GmbH].
Figure \ref{tests} shows representative results obtained from uni-axial compression and tension tests, in the neighborhood of the unstressed state.
We carried out several tests in each regime, on specimens with various sizes, and found that in the linear \emph{tensile} region, the slope was approximatively 0.45 MPa (notice that the silicone is about 20 times softer in tension than the NR70 of Bechir et al.), while in the \emph{compressive} region it  was approximatively 0.25 MPa.
This suggest that in contrast to rubber, silicone is stiffer in tension than in compression, with 
\be
E^+ / E^- \simeq 1.8.
\en

Of course, we must acknowledge that external factors, other than bimodularity, can explain that $E^+/E^- \ne 1$, most importantly, those due to experimental error and protocol. 
Indeed uniaxial compressive tests such as those presented above are harder to implement than tensile tests, because a perfect lubrication between the platens and the sample is required in order to ensure homogeneous deformations. 
Nonetheless we note that there exists standardized protocols for compressive tests of rubbers (see ISO 7743 \cite{ISO}, ASTM D575 \cite{ASTM}, and also Gent \cite{Gent}).
It seems thus perfectly legitimate to investigate what would be the consequences of bimodularity for elastomers.

Finally we remark that Soltz and Ateshian \cite{Cart} used bimodularity to model the tension-compression nonlinearity of articular cartilage. 
There, the mean values obtained from 9 specimens give $E^+ / E^- = 12.75/0.6 = 21.25$.
Similarly, Bertoldi et al. \cite{BeBD08} recently showed that nacre (mother-of-pearl) must be considered orthotropic and bimodular in order to interpret correctly the available experimental data. They estimate that for nacre, $0.5 < E^+/E^- < 0.8$. 
They also carry out calculations for the plane strain bending of a orthotropic, bimodular beam, within the theory of linear elasticity. 
In what follows here, we focus on the large bending of isotropic, incompressible, bimodular beams in non-linear elasticity. 


\section{Bending}
\label{bending_section}



\subsection{Bending deformation}
\label{Bending_deformation}


The fundamental assumption introduced by Rivlin \cite{Rivl49} to model the non-linear flexure of an incompressible beam is that a beam of length $L$ and thickness $2A$ is deformed under applied terminal moments into a circular, annular sector. 
For definiteness, assume that the faces $X=\pm A$ are deformed into the inner and outer faces of radii $r_a$ and $r_b$, respectively, of the annular sector and that the faces $Y=\pm L/2$ are deformed into the faces $\theta=\pm \alpha$, where $\alpha$ is a $\emph{specified}$ constant. 
The \emph{turning angle} $\alpha$, is restricted to lie in the range 
\begin{equation}
0 \leq \alpha \le \pi,
\end{equation}
which only allows a beam to be bent into at most a circular annulus. 

Adopting a semi-inverse approach, Rivlin \cite{Rivl49} shows that the following non-ho\-mo\-ge\-ne\-ous deformation field is a solution to the equilibrium equations of non-linear incompressible elasticity:
\begin{equation} \label{rivs}
r = \sqrt{2 (L/\alpha)X + D}, \qquad \theta = \alpha Y/L, \qquad z = Z,
\end{equation}
where ($X, Y, Z$) and ($r, \theta, z$) denote the Cartesian and cylindrical polar coordinates of a typical particle before and after deformation, respectively, and $D$ is a constant.
Hence, the inner and outer radii of the deformed curved surfaces are determined by
\begin{equation}
r _{a,b} = \sqrt{D \mp 2 (L/\alpha)A }.
\end{equation}
Adding and subtracting these equations then yields
\begin{equation}\label{r_ab}
D = (r_a^2 + r_b^2)/2, \qquad
r_b^2 - r_a^2 = 4 A L / \alpha.
\end{equation}

The corresponding deformation gradient tensor, $\vec{F}$, is given by
\begin{equation} \label{F}
\vec{F} = \text{diag}\left(\lambda_1, \lambda_2, \lambda_3\right)
= \text{diag}\left(\lambda, \lambda^{-1}, 1\right), \quad \text{where}
\quad \lambda = L/(\alpha r),
\end{equation}
denoting the principal stretches by $\lambda_1$, $\lambda_2$, $\lambda_3$.
Here $\lambda_2 = \lambda^{-1}$ is the stretch experienced by circumferential line elements: when $\lambda_2 > 1$ ($\lambda<1$), the line elements are extended by the bending; when $\lambda_2<1$ ($\lambda>1$), they are contracted.
Also, $\lambda_3=1$ at all times, showing that Rivlin's solution \eqref{rivs} is a plane strain deformation. 
Note that $\lambda_1 \lambda_2 \lambda_3 = 1$, showing that  the deformation respects the internal constraint of incompressibility. 
Finally, it is clear that $\vec{F}$ corresponds to a non-homogeneous deformation because the radial stretch $\lambda_1$ varies as the inverse of $r$.
In particular, it takes the following values 
\begin{equation} \label{l_ab}
\lambda_a = L/(\alpha r_a), \qquad \lambda_b = L/(\alpha r_b),
\end{equation}
on the inner and outer faces of the bent beam, respectively.
These notations allow us to rewrite \eqref{r_ab}$_2$ in non-dimensional form as
\be \label{lambdas1}
\lambda_b^{-2} - \lambda_a^{-2} = 2\epsilon,
\en
where $\epsilon$ is the product of the beam aspect ratio by the turning angle \cite{DGM}:
\be \label{epsilon}
\epsilon = (2A/L)\alpha.
\en


\subsection{Homogeneous (unimodular) beams}
\label{Bending_stress}


For homogeneous, incompressible, elastic materials, the corresponding principal Cauchy stresses are
\begin{equation}
T_{rr} = -p + \lambda_1 W_{,1} , \qquad T_{\theta \theta} = -p + \lambda_2 W_{,2},
\end{equation}
where $p$ is an arbitrary scalar field,  $W =   W(\lambda_1, \lambda_2, \lambda_3)$ is the strain-energy function and the comma subscript denotes partial differentiation with respect to the appropriate principal stretch. 
The equations of equilibrium determine $p$ as
\begin{equation}
p = \int \left(\lambda_1 W_{,1} - \lambda_2 W_{,2}\right) r^{-1} \text{d}r
 + \lambda_1 W_{,1} + K,
\end{equation}
where $K$ is an arbitrary constant. 
It therefore follows immediately that
\begin{equation} \label{Trr}
T_{rr} =  \int \left(\lambda_1 W_{,1} - \lambda_2 W_{,2}\right) r^{-1} \text{d}r
 + K,
 \qquad
T_{\theta \theta} =  T_{rr} +  \lambda_2 W_{,2} - \lambda_1 W_{,1}.
\end{equation}

Now define the function $\widetilde W(\lambda)$ as
\begin{equation}
\widetilde W (\lambda) = W(\lambda, \lambda^{-1}, 1),
\end{equation}
which is assumed henceforth to be a convex function. 
Then 
\begin{equation} \label{zerostress}
\lambda \widetilde{W}' = \lambda_1 W_{,1} - \lambda_2 W_{,2},
\end{equation} 
where the prime denotes differentiation.
The stress components can then be written simply as functions of $\lambda$  as
\begin{equation} \label{T}
T_{rr} = \widetilde W + K, \qquad T_{\theta \theta} = T_{rr} - \lambda \widetilde{W}'.
\end{equation}
The curved surfaces of the bent beam are assumed to be free of traction.
This assumption then yields
\begin{equation} \label{K}
K = -\widetilde W (\lambda_a), \qquad \widetilde W (\lambda_b) = \widetilde W(\lambda_a).
\end{equation}

Only \emph{isotropic} materials are considered in this paper. 
For these materials, $W(\lambda_1, \lambda_2, 1) = W(\lambda_2, \lambda_1, 1)$ ,  or equivalently, 
\begin{equation} \label{iso}
\widetilde W(\lambda) = \widetilde W(\lambda^{-1}).
\end{equation}
Then \eqref{K}$_2$ yields
\be \label{sym}
\widetilde W(\lambda_b) = \widetilde W(\lambda_a) = 
\widetilde W(\lambda_b^{-1}) = \widetilde W(\lambda_a^{-1}).
\en
There are two obvious solutions to these equations: the first is $\lambda_a = \lambda_b$, which by \eqref{l_ab} gives $r_a = r_b$, a physically unacceptable solution; 
the second is $\lambda_a = \lambda_b^{-1}$, which  gives
\begin{equation} \label{sufficient}
\lambda_a \lambda_b = 1, \qquad
\text{or} \qquad
\alpha^2 r_a r_b = L^2.
\end{equation}
In that latter case,  \eqref{l_ab} yields a quadratic equation for  $\lambda_a^2$, with the following unique physically acceptable solution:
\begin{equation} \label{r_a^2}
\lambda_a = \sqrt{\epsilon + \sqrt{\epsilon^2+1}}, \qquad
\text{or} \qquad
r_a^2 = \dfrac{L}{\alpha} \left( \sqrt{4 A^2 + \dfrac{L^2}{\alpha^2}} - 2A \right),
\end{equation}
which completely determines the deformed configuration. Unusually, it is \emph{independent of the form of the strain-energy function} (provided that \eqref{sufficient}, which is sufficient for \eqref{sym} to be satisfied, is also necessary). 
From the specialization to isotropic materials follows also immediately from \eqref{zerostress} that
\begin{equation} \label{zeroderiv}
\widetilde{W}'(1)=0.
\end{equation}

Hereafter we consider the boundary value problem where equal and opposite moments are applied to the ends of the beam at $Y = \pm L/2$, whilst the inner and outer curved faces of the beam are traction-free.


\subsection{Bimodular beams}
\label{Neutral_axis}


The \emph{neutral axis} in flexure is the circumferential material line (originally parallel to the major axis of the beam in the reference configuration) with unchanged length, see Figure \ref{fig_bending}.
The neutral axis is therefore given by the radius $r_n$ in the deformed configuration such that
\begin{equation}
r_n=\frac{L}{\alpha},
\end{equation}
assuming that 
\begin{equation}
r_a \leq r_n \leq r_b.
\end{equation}
The neutral axis is the natural, \emph{intuitive}, boundary in flexure between the regions in ``tension'' and ``compression'', defined by $r>r_n$ and $r<r_n$, respectively. 

Assume now that a material behaves differently in these two regions. 
The turning angle $\alpha$ is assumed to be constant across these two regions. 
Assuming also that the radial deformation field is continuous across $r_n$ then means that both the deformation field and the deformation gradient tensor have the same form in both regions. 
Denote quantities associated with the tensile and compressive regions by the superscripts `+' and `-', respectively. 
It follows from \eqref{T} that the radial stress therefore has the form
\begin{equation}
T_{rr}^{+}= \widetilde W^{+}(\lambda)+K^+, \qquad 
T_{rr}^{-} = \widetilde W^{-}(\lambda)+K^-,
\end{equation}
where $K^+, K^-$ are constants and $\lambda$ is defined by \eqref{F}. 
Satisfying the stress free boundary conditions on the upper and lower curved surfaces, and solving for $K^+$, $K^-$ then yields
\begin{equation}
T^+_{rr} = \widetilde W^+(\lambda) - \widetilde W^+(\lambda_b), \qquad 
T^-_{rr} = \widetilde W^-(\lambda) - \widetilde W^-(\lambda_a).
\end{equation}
Along the neutral axis then 
\begin{equation}
T^+_{rr}(r_n)= \widetilde W^+(1) - \widetilde W^+(\lambda_b), \qquad 
T^-_{rr}(r_n)= \widetilde W^-(1) - \widetilde W^-(\lambda_a).
\end{equation}
The usual assumption of zero strain energy in the reference configuration yields
\begin{equation}
\widetilde W^+(1)= \widetilde W^-(1)=0,
\end{equation}
and so we obtain the following relation for \emph{the normal stress difference across the neutral axis}:
\begin{equation} \label{delta}
\Delta T_{rr}\equiv T_{rr}^+(r_n)-T_{rr}^-(r_n)= \widetilde W^-(\lambda_a)- \widetilde W^+(\lambda_b).
\end{equation}
It follows from \eqref{Trr} and \eqref{zeroderiv} that \emph{the difference in the hoop stress is exactly the same}:
\begin{equation} \label{delta2}
\Delta T_{\theta \theta}\equiv T_{\theta \theta}^+(r_n)-T_{\theta \theta}^-(r_n)= \widetilde W^-(\lambda_a)- \widetilde W^+(\lambda_b).
\end{equation}
We call $\Delta T$ this stress difference: $\Delta T = \Delta T_{rr} = \Delta T_{\theta \theta}$.

Consider now that there is \emph{no stress discontinuity across the neutral radius}. 
Then $\Delta T = 0$ and 
\begin{equation}
\widetilde W^-(\lambda_a)= \widetilde W^+(\lambda_b).
\end{equation}
Recalling \eqref{lambdas1}, this equation can be rewritten as the defining equation for $\lambda_a$ as follows:
\begin{equation} \label{deflam}
\widetilde W^-(\lambda_a)= \widetilde W^+(\lambda_a(1+2\epsilon\lambda_a^2)^{-\frac{1}{2}}).
\end{equation}
With $\lambda_a$, necessarily $>1$, determined in this way, the deformed configuration is completely determined. 
Note that, in contrast to the standard flexure problem solved by Rivlin (see \eqref{r_a^2}), the deformed configuration depends here on the form of the strain-energy function. 
One such form is considered in the next section.


\section{An example}
\label{example_section}


In this section we solve the flexure problem in the case where the solid is modeled by the Mooney-Rivlin strain energy density:
\be
W = \frac{\mu}{2}(1+f)(\lambda_1^2 + \lambda_2^2 + \lambda_3^2-3)
+  \frac{\mu}{2}(1-f)(\lambda_1^2\lambda_2^2 + \lambda_2^2\lambda_3^2 + \lambda_3^2\lambda_1^2-3),
\en 
where $\mu$ and $f$ are constants, or equivalently \cite{GoVD08}, by the third-order expansion of $W$ in the weakly non-linear elasticity approximation \cite{Ogde74, HaIZ04}:
\begin{equation} \label{3rd}
W = \mu \; \text{tr}(\vec{E}^2) + \frac{\mathcal{A}}{3} \text{tr}(\vec{E}^3),
\end{equation}
where $\mu$ is the shear of modulus of second-order elasticity, $\mathcal{A}$ is a third-order non-linear Landau constant, and  $\vec{E}$ is the Green-Lagrange strain tensor (with eigenvalues $(\lambda_i^2-1)/2$).

In both cases we find that
\begin{equation} \label{neo}
\widetilde W (\lambda) = \frac{\mu}{2}\left(\lambda^2+\lambda^{-2}-2\right).
\end{equation}
Then \eqref{deflam} reduces to the following cubic:
\begin{equation} \label{defx}
2\epsilon  x^3+\left[1-4\epsilon-\hat{\mu}\left(1-2\epsilon\right)^2\right]x^2 - 2\left[1 - \epsilon - \hat{\mu}\left(1-2\epsilon\right)\right]x + 1-\hat{\mu}=0,
\end{equation}
where $x\equiv\lambda_a^2$, $\epsilon$ is defined in \eqref{epsilon}, and $\hat{\mu}\equiv \mu^+ /\mu^-$ is the ratio of the shear modulus in the region of tension by the shear modulus in the region of compression. 

\begin{figure}[ht]
\label{fig_lambda_a}
\begin{center}
\includegraphics[width=11cm]{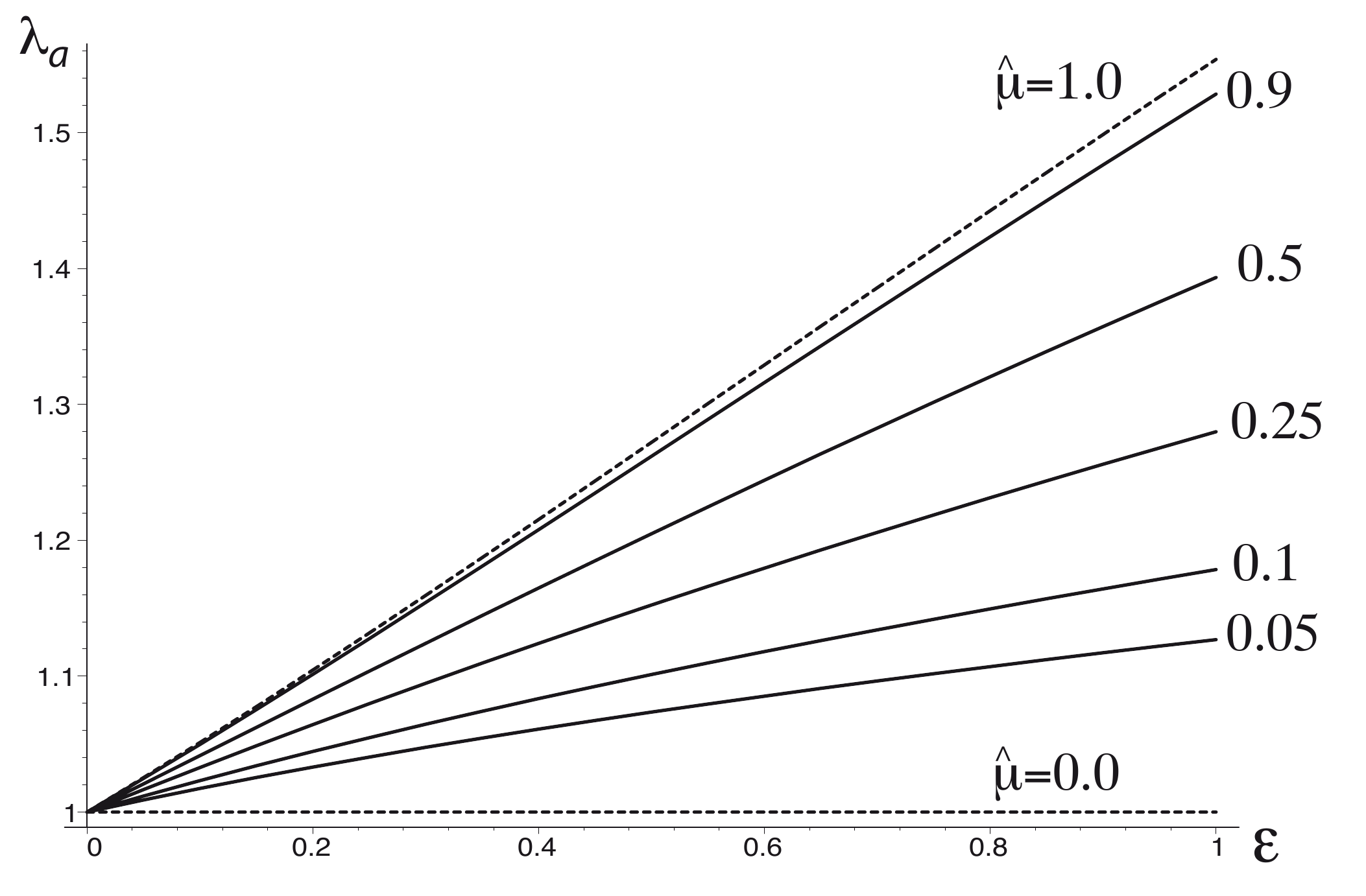}\end{center}
\caption{Variations of $\lambda_a$, the radial stretch ratio on the inner curved face, with $\epsilon$, the product of the bar aspect ratio by the turning angle, in the case of a bimodular Mooney-Rivlin solid. 
The ratio of the tension shear modulus to the compression shear modulus takes the values: $\hat \mu = 0.0$ (horizontal dotted line), $1/20$, $1/10$, $1/4$, $1/2$, $1/1.1$ (full lines), and 1.0 (other dotted line).}
\end{figure}

Focusing now on bimodular solids which are stiffer in compression than in tension, we take $0<\hat \mu <1$.
Calling $f(x)$ the cubic on the left hand side of \eqref{defx}, we find that $f(1)=-4 \hat \mu \epsilon^2 <0$ and $f(\infty) = 1-\hat \mu>0$, ensuring that there always exists a relevant root for $\lambda_a$.
Setting $\hat{\mu}=1$ recovers the case where there is no difference in material properties in tension and compression, and \eqref{defx} then yields \eqref{r_a^2}.
Letting $\hat{\mu}\rightarrow0$ in \eqref{defx}, which physically corresponds to the case where $\mu^{+}<<\mu^{-}$, yields $\lambda_a=1$: in other words, an infinitely stiff bar cannot be bent, as expected intuitively.
Figure \ref{fig_lambda_a} displays several graphs of $\lambda_a$ as a function of $\epsilon$, for different values of the parameter $\hat \mu$. 
It is clear that the effect of the tension-compression difference becomes more pronounced with increasing angle and/or aspect ratio.



\section{Possibility of a stress discontinuity}
\label{Stress_discontinuity}


Finally, the possibility of a stress discontinuity across the neutral axis is considered. The size of the stress discontinuity cannot be determined without a further assumption. It will be assumed here that the deformed configuration for the bimodular material is the same as that in the classical problem of flexure,

As shown in Section \ref{Bending_stress}, the deformation of the classical flexure problem  is independent of the form of the strain-energy function and therefore, in particular, independent of which of the tension/compression forms is to assumed. 
For the classical problem,
\begin{equation}
\lambda_b=\lambda^{-1}_{a},
\end{equation}
according to \eqref{r_ab}
and, from \eqref{delta}, the stress discontinuity in the bimodular beam therefore has the value
\begin{equation} \label{delta1}
\Delta T= \widetilde W^-(\lambda_a)- \widetilde W^+(\lambda_{a}^{-1}).
\end{equation}
Noting \eqref{iso}, this  can be simplified as  
\begin{equation} \label{delta4}
\Delta T= \widetilde W^-(\lambda_a)- \widetilde W^+(\lambda_a).
\end{equation}

Now the material is isotropic, and without loss of generality, the strain-energy function can be assumed to have the form
\begin{equation}
\widetilde{W}(\lambda)=\widetilde{W}(I),\qquad \text{where}
\qquad
I\equiv\lambda^{2}+\lambda^{-2}-2,
\end{equation}
because in bending, the first and second principal invariants of the Cauchy-Green strain tensor are both equal to $I$.
The problem of flexure is a typical example of a deformation where the strain, quantified here by $I$, is small but with possibly moderate or large deformations. Therefore the general strain-energy function $\widetilde{W}(\lambda)$ can be closely approximated by the linearisation of $\widetilde{W}(I)$, i.e., by the form  \eqref{neo} of the strain-energy function. 
It follows that $\Delta T$ can be closely approximated in general by
\begin{equation} \label{delta3}
\Delta T= \left(\mu^{-}-\mu^{+}\right)\left(\lambda^{2}_{a}+\lambda^{-2}_{a}-2\right)
= 2 \left(\mu^{-}-\mu^{+}\right)\left(\sqrt{\epsilon^2+1} - 1 \right).
\end{equation}

Hence, in general, it can be seen that the stress discontinuity cannot be large because the difference in the shear moduli is multiplied by a small term. 
We would like to propose this stress difference as a possible cause of \emph{damage} in the flexure of hyperelastic  materials. 
We note that simply flexing a hyperelastic bar is not likely to cause much damage but, over time, flexing and unflexing the same bar is likely to cause the accumulation of small amounts of damage due to the stress discontinuity, with inevitable consequences for the integrity of the component.


\section*{Acknowledgments}


This work was supported by a Marie Curie Fellowship for Career Development awarded by the Seventh Framework Programme of European Commission (for the first author) and  by a Postgraduate Research Scholarship awarded by the Irish Research Council for Science, Engineering and Technology (for the third author).

We are greatly indebted to Hocine Bechir for providing us with the experimental data of \cite{Bechir}.



\end{document}